\documentclass[%
 reprint,
superscriptaddress,
 amsmath,amssymb,
 aps,
]{revtex4-1}

\usepackage{graphicx}
\usepackage{dcolumn}
\usepackage{bm}
\usepackage{physics}
\usepackage{mathrsfs}

\begin{document}

\title{Optimized protocol for twin-field quantum key distribution}

\author{Rong Wang}
\author{Zhen-Qiang Yin}
\email{yinzq@ustc.edu.cn}
\author{Feng-Yu Lu}
\author{Shuang Wang}
\author{Wei Chen}
\affiliation{CAS Key Laboratory of Quantum Information, CAS Center for Excellence in Quantum Information and Quantum Physics, University of Science and Technology of China, Hefei 230026, China}
\affiliation{State Key Laboratory of Cryptology, P. O. Box 5159, Beijing 100878, P. R. China}
\author{Chun-Mei Zhang}
\affiliation{Institute of Quantum Information and Technology, Nanjing University of Posts and Telecommunications, Nanjing 210003, China}
\author{Wei Huang}
\author{Bing-Jie Xu}
\affiliation{Science and Technology on Communication Security Laboratory, Institute of Southwestern Communication, Chengdu, Sichuan 610041, China}
\author{Guang-Can Guo}
\author{Zheng-Fu Han}
\affiliation{CAS Key Laboratory of Quantum Information, CAS Center for Excellence in Quantum Information and Quantum Physics, University of Science and Technology of China, Hefei 230026, China}
\affiliation{State Key Laboratory of Cryptology, P. O. Box 5159, Beijing 100878, P. R. China}

\begin{abstract}
Twin-field quantum key distribution (TF-QKD) and its variant protocols are highly attractive due to the advantage of overcoming the rate-loss limit for secret key rates of point-to-point QKD protocols. For variations of TF-QKD, the key point to ensure security is switching randomly between a code mode and a test mode. Among all TF-QKD protocols, their code modes are very different, e.g. modulating continuous phases, modulating only two opposite phases, and sending or not sending signal pulses. Here we show that, by discretizing the number of global phases in the code mode, we can give a unified view on the first two types of TF-QKD protocols, and demonstrate that increasing the number of discrete phases extends the achievable distance, and as a trade-off, lowers the secret key rate at short distances due to the phase post-selection. 
\end{abstract}

\pacs{Valid PACS appear here}
\maketitle


\noindent{\bf INTRODUCTION}

\noindent Quantum key distribution (QKD) \cite{BB84,Ekert1991quantum} provides two distant parties (Alice and Bob) with a secure random bit string against any eavesdropper (Eve) guaranteed by the law of quantum mechanics. During last three decades, QKD has rapidly developed both in theory and experiment \cite{Inoue2002DPS,Lo2012MDI,gobby2004quantum,Jouguet2013Experimental,Wang2015RFIMDI}, and it is on the way to a wide range of implementation. Among all QKD experiments before, without quantum repeaters, the maximum key rates are bounded with respect to the channel transmittance $\eta$, defined as the probability for an effective detector click caused by a transmitted photon. So, one of the crucial tasks for the theorists is to find the maximum key rate achievable under ideal implementation (based on perfect single-photon sources, pure-loss channels, perfect detectors, perfect post-processing and so on). With the aim of finding an upper bound of secret key rate, the theorists have provided several answers \cite{pirandola2009direct,takeoka2014fundamental,pirandola2017fundamental}. The recent work has provided the fundamental limit called Pirandola-Laurenza-Ottaviani-Banchi (PLOB) bound\cite{pirandola2017fundamental}, which establishes that the secret key rate without quantum repeaters must satisfy $R\le -log(1-\eta)$.

Remarkably, the twin-field (TF) QKD protocol, proposed by Lucamarini et al. \cite{lucamarini2018overcoming}, is capable of overcoming this PLOB bound with some restrictions on Eve's strategies, which is mainly attributed to the single-photon interferometric measurement at the third untrusted party Eve. In other words, a single photon came from either Alice or Bob interferes at Eve's beam splitter and clicks the detector,  which means that generating a secret key bears a unilateral transmission loss. Because of this dramatic breakthrough, a variant of TF-QKD protocols have been proposed consequentially \cite{ma2018phase,wang2018twin,cui2019twin,curty2019simple,lin2018simple,tamaki2018information} and some protocols have been demonstrated experimentally \cite{minder2019experimental,wang2019beating,liu2019experimental,zhong2019proof,chen2020sending}. For variant TF-QKD protocols, the key idea to ensure the security is switching probabilistically between a code mode and a test mode, where the former is for key generation, and the latter is for parameter estimation\cite{tamaki2018information}. Among all TF-QKD protocols, their code modes are very different, e.g. modulating continuous phases \cite{lucamarini2018overcoming,ma2018phase}, modulating only two opposite phases \cite{cui2019twin,curty2019simple,lin2018simple}, and sending or not sending signal pulses\cite{wang2018twin}. The code modes of the first two kinds are similar in some sense. Intuitively, they may be explained by a unified view.

Interestingly, by discretizing the global phases of Alice and Bob's emitted pulses in the code mode, we can give a unified view on two kinds of TF-QKD protocols \cite{lucamarini2018overcoming,ma2018phase,cui2019twin,curty2019simple,lin2018simple}. Specifically, Alice and Bob encode classical bit $0$, $1$ into phases $0$, $\pi$ of a coherent state, respectively, then randomize them by adding a phase chosen randomly $0, \pi/M, 2\pi/M, \ldots, (M-1)\pi/M$.  According to whether or not perform phase post-selection in the test mode, we introduce two protocols. To prove their security, we establish a universal framework against collective attacks, which can be extended to robust against coherent attacks \cite{lu2019practical} with the technique in \cite{christandl2009postselection}. The security analysis indicates that increasing the number of discrete phases can extend the achievable distance, but lower the secret key rate at short distances due to the phase post-selection. Furthermore, simulation results show that a small number of random phases (say M=2) may be the best choice for practical implementations.


\hfill

\noindent{\bf RESULTS}

\noindent We firstly describe details of our proposed TF-QKD protocols that have discrete phase randomization in the code mode, and the schematic setup is shown in Fig 1.

\begin{figure}[htbp]
\centering
\includegraphics[width=\linewidth]{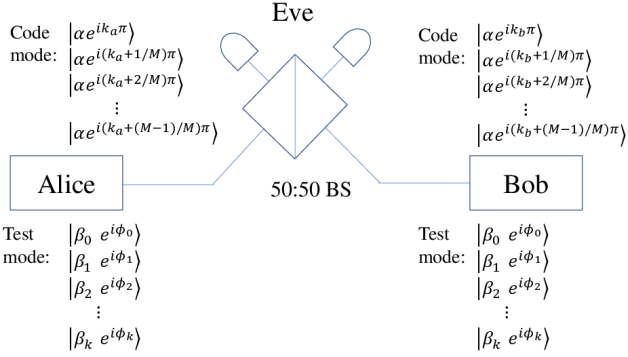}
\caption{ Schematic setup of our Twin-field quantum key distribution protocols: In each trial, Alice and Bob randomly choose code mode and test mode and send their quantum states to the untrusted receiver Eve. If a code mode is selected, Alice (Bob) prepares coherent state chosen from $\{\ket{\alpha e^{ik_{a(b)}\pi}}, \ket{\alpha e^{i(k_{a(b)}+\frac{1}{M})\pi}}, \ldots, \ket{\alpha e^{i(k_{a(b)}+\frac{M-1}{M})\pi}}\}$. If a test mode is selected, Alice (Bob) prepares coherent state chosen from $\{\ket{\beta_{0}e^{i\phi_{0}}}, \ket{\beta_{1}e^{i\phi_{1}}},\ldots,\ket{\beta_{k}e^{i\phi_{k}}}\}$. After interference at beam splitter (BS) and detector click on Eve's side, she announces the outcome. More detailed explanation can be found in protocol descriptions.}.
\label{fig:false-color}
\end{figure}

{\it Protocol} \uppercase\expandafter{\romannumeral1}

Step 1. Alice and Bob randomly choose code mode or test mode in each trial.

Step 2. If a code mode is selected, Alice (Bob) randomly generates a key bit $k_{a}$ ($k_{b}$) and a random number $x$ ($y$) and then prepares the coherent state $\ket{\alpha e^{i(k_{a}+\frac{x}{M})\pi}}$ ($\ket{\alpha e^{i(k_{b}+\frac{y}{M})\pi}}$), where $x,y\in{\{0,1,2,\ldots, M-1\}}$. If a test mode is selected, Alice (Bob) generates a random phase $\phi_{a}\in{[0,2\pi)}$ ($\phi_{b}\in{[0,2\pi)}$) and emits coherent state $\ket{\beta_{a}e^{i\phi_{a}}}$ ($\ket{\beta_{b}e^{i\phi_{b}}}$), where $\beta_{a}$ ($\beta_{b}$) is randomly chosen from a pre-decided set.

Step 3. Alice and Bob send their quantum states to the untrusted receiver Eve. For each trial, only three outcomes are acceptable, which are "Only detector $L$ clicks", "Only detector $R$ clicks" and "No detectors click", and Eve announces one of these outcomes. Note that the outcome "Both detectors $L$ and $R$ click" is considered as "No detectors click".

Step 4. Alice and Bob repeat the above steps many times. For the successful detection outcomes (only detector L or R clicks), Alice and Bob publicly announce which trials are the code modes and which trials are the test modes. For each successful trial in the code mode, Alice and Bob announce their $x$ and $y$, and keep $k_{a}$, $k_{b}$ as their raw key if $x=y$. Moreover, Bob should flip his key $k_{b}$ if Eve announces "Only detector $R$ clicks". 

Step 5. For each trial that both Alice and Bob select test mode, Alice and Bob announce ${\beta_{a}}$ with random phase $\phi_{a}$ and ${\beta_{b}}$ with random phase $\phi_{b}$, and only keep the trial that $\beta_{a}=\beta_{b}$ and $|\phi_{a}-\phi_{b}|=0$ or $\pi$.

Step 6. Alice and Bob perform information reconciliation and privacy amplification to extract the final secure keys.

For the simplicity in experiments, we can remove post-selection in the test mode, and the simplified protocol runs as follows.

{\it Protocol} \uppercase\expandafter{\romannumeral2}

Step 1. Same as Protocol \uppercase\expandafter{\romannumeral1}

Step 2. Same as Protocol \uppercase\expandafter{\romannumeral1}

Step 3. Alice and Bob send their quantum states to the untrusted receiver Eve. For each trial, only three outcomes are acceptable, which are "Only detector $L$ clicks", "Only detector $R$ clicks" and "No detectors click". Note that, the outcome "Both detectors $L$ and $R$ click" is considered as "No detectors click" in the code mode, and is considered as only detector $L$ or $R$ clicks with equal probability in the test mode. Consequentially, Eve announces one of these outcomes. 

Step 4. Alice and Bob repeat the above steps many times. For the successful detection outcomes (only detector $L$ or $R$ clicks),  Alice and Bob publicly announce which trials are the code modes and which trials are the test modes. For each successful trial in the code mode, Alice and Bob announce their $x$ and $y$, and keep $k_{a}$, $k_{b}$ as their raw key if $x=y$. Moreover, Bob should flip his key $k_{b}$ if Eve announces "Only detector $R$ clicks". 

Step 5. For each trial that both Alice and Bob select the test mode, the yield $Y_{l,k}$, probability of Eve announcing the successful outcome provided Alice emits $l$-photon state and Bob emits $k$-photon state, can be estimated. 

Step 6. Same as Protocol \uppercase\expandafter{\romannumeral1}

\begin{table}[htbp]
\centering
\caption{Parameters}
\label{table1}
\renewcommand{\arraystretch}{1.2}
\setlength{\tabcolsep}{9mm}{
\begin{tabular}{ll}

\hline
Parameters & Values \\
\hline
Dark count rate $d$ & $8\times10^{-8}$\\
Error correction efficiency $f$ & 1.15 \\
Detector efficiency $\eta_d$ & 14.5\%\\
Misalignment error $e_{mis}$ &  1.5\%\\
\hline

\end{tabular}}

\end{table}

Our security proof is based on Devetak-Winter’s bound \cite{devetak2005distillation}, concretely, bounding the information leakage $I_{AE}$. Thus, the secret key rate is given by
\begin{equation}
R\ge \frac{1}{M}Q(1-fH(e)-I_{AE}),
\end{equation}
where $Q$ is the counting rate, $1/M$ is the shifting factor, $e$ is the error rate, and $f$ is the error correction efficiency. By applying infinite decoy states \cite{Hwang2003Decoy,Lo2005Decoy,Wang2005Decoy,Ma2005Decoy} in the test mode, we can simulate the performance of our two protocols with different $M$. The simulation parameters are given in Table \uppercase\expandafter{\romannumeral1}. For Protocol \uppercase\expandafter{\romannumeral1}, we present the numerical simulations of secret key rate in Fig 2 and the maximal channel loss in Table \uppercase\expandafter{\romannumeral2}. If we remove the sifting efficiency, the limitary channel loss with $M\to \infty$ is $81.5$ dB as shown in Table \uppercase\expandafter{\romannumeral2}. According to Fig 2 and Table \uppercase\expandafter{\romannumeral2}, it's sufficient to apply TF-QKD with $M=2$ which almost reaches the theoretical limit channel loss. Analogously, for Protocol \uppercase\expandafter{\romannumeral2}, we get simulation results comparable to those of Protocol \uppercase\expandafter{\romannumeral1}, and we show the secret key rate in Fig 3 and the theoretical limit channel loss in Table \uppercase\expandafter{\romannumeral3}. When removing  the sifting factor, the maximal channel loss of Protocol \uppercase\expandafter{\romannumeral2} with$M\to \infty$ is $75.8$ dB.

\begin{figure}[htbp]
\centering
\includegraphics[width=\linewidth]{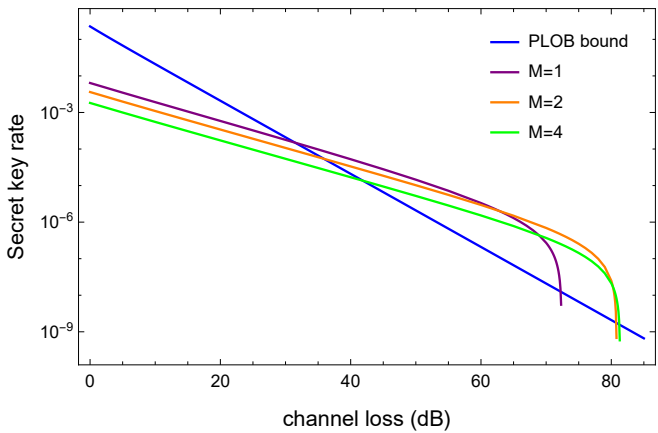}
\caption{Secret key rate R versus channel loss for Protocol \uppercase\expandafter{\romannumeral1}: The curves represent the secure key rate of twin-field quantum key distribution protocol for $M=1$, $M=2$, $M=4$ ($M$ is the number of random phases) and the Pirandola-Laurenza-Ottaviani-Banchi (PLOB) bound respectively. We do not show the case of $M\to \infty$ because the key rate tends to $0$}.
\label{fig:false-color}
\end{figure}

\begin{table}[htbp]
\centering
\caption{The maximal channel loss for Protocol \uppercase\expandafter{\romannumeral1} with different $M$.}
\label{table1}
\renewcommand{\arraystretch}{1.1}
\setlength{\tabcolsep}{5mm}{
\begin{tabular}{c|c}

$M$ & The maximal channel loss (dB) \\
\hline
1 & 72.3  \\
2 & 80.8 \\
4 & 81.3 \\
$\infty$ & 81.5 \\

\end{tabular}}

\end{table}

When we compare Protocol \uppercase\expandafter{\romannumeral1} with Protocol \uppercase\expandafter{\romannumeral2}, the latter one does not require post-selection in the test mode, as a trade-off, the maximal channel loss will be lower. Here, we consider the relationship with several varietal TF-QKD protocols \cite{ma2018phase,cui2019twin,curty2019simple,lin2018simple}. When $M\to \infty$, Protocol \uppercase\expandafter{\romannumeral1} is exactly the PM-QKD \cite{ma2018phase} if we relax the post-selection condition $|\phi_{a}-\phi_{b}|=0$ or $\pi$ and add a corresponding sifting factor. When $M=1$, Protocol \uppercase\expandafter{\romannumeral2} is the same as \cite{cui2019twin,curty2019simple,lin2018simple} in the code mode, the difference is the way to estimate the information leakage or the "phase error". To some extent, our proposed TF-QKD protocols with discrete phase randomization in code mode cover the four varietal TF-QKD protocols above.

\begin{figure}[htbp]
\centering
\includegraphics[width=\linewidth]{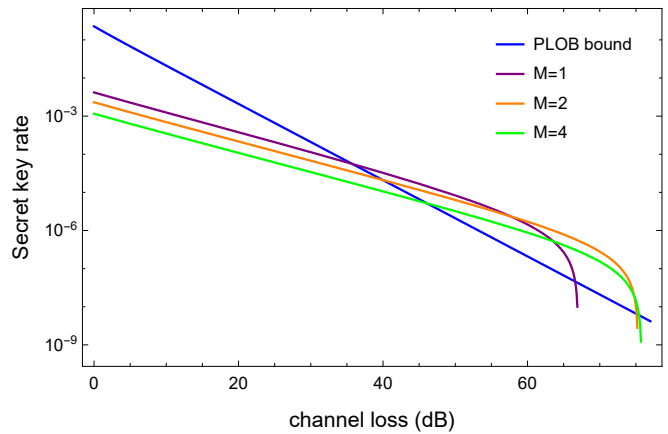}
\caption{Secret key rate R versus channel loss for Protocol \uppercase\expandafter{\romannumeral2}: The curves represent the secure key rate of twin-field quantum key distribution protocol for $M=1$, $M=2$, $M=4$ ($M$ is the number of random phases) and the Pirandola-Laurenza-Ottaviani-Banchi (PLOB) bound respectively. We do not show the case of $M\to \infty$ because the key rate tends to $0$}
\label{fig:false-color}
\end{figure}

\begin{table}[htbp]
\centering
\caption{The maximal channel loss with Protocol \uppercase\expandafter{\romannumeral2} different $M$.}
\label{table1}
\renewcommand{\arraystretch}{1.1}
\setlength{\tabcolsep}{5mm}{
\begin{tabular}{c|c}

$M$ & The maximal channel loss (dB) \\
\hline
1 & 67.0  \\
2 & 75.3 \\
4 & 75.8 \\
$\infty$ & 75.8 \\

\end{tabular}}

\end{table}

\hfill

\noindent{\bf DISCUSSION}

In summary, we have introduced a variant TF-QKD with discrete phase randomization in the code mode and proven its security in asymptotic scenarios. Our protocol can be viewed as a generalization of the four varietal TF-QKD protocols \cite{ma2018phase,cui2019twin,curty2019simple,lin2018simple} to some extent. The security proof discloses that the transmission distance becomes longer with $M$ exponentially increasing, as a trade-off, the secret key rate is lower at short distances. As a result, the transmission distance reaches a limitation when $M$ tends to infinity. Numerical simulations show that it's sufficient to apply TF-QKD with $M=2$, for it almost reaches the limitary transmission distance at the cost of about half of secret key rate, compared with the case of $M=1$, at short distance. Furthermore, post-selection in the test mode is not convenient in experiment, thus, we remove it to make experiments simpler in a modified protocol. We find that the removal of post-selection in the test mode has very limited influence on the secret key rate and achievable distance. Our findings expect TF-QKD can be run with optimal phase randomization actively, i.e. at short distance one can simply bypass phase randomization, while a phase randomization of $0$ or $\pi/2$ is sufficient at the long distance case. 

During the preparation of this paper, we find that Primaatmaja et al. \cite{primaatmaja2019versatile} proposed an open question that if coding phase in TF-QKD under different bases can improve secret key rate significantly. Their open question is answered by our finding that $M=2$ is almost optimal  in some sense.

\hfill

\noindent{\bf METHODS}

\noindent{\bf Security proof}

\noindent Here, we present security proof Protocol \uppercase\expandafter{\romannumeral1}. Firstly, we analyze the composite states shared by Alice and Bob when they both select the test mode. In the case of $\beta_{a}=\beta_{b}=\beta$ and $\phi_{a}=\phi_{b}=\phi$, the composite state of Alice and Bob can be written as
\begin{equation}
\label{constext}
\begin{aligned}
\rho_{AB}&=\frac{1}{2\pi}\int_{0}^{2\pi}d\phi\ket{\beta e^{i\phi}}\ket{\beta e^{i\phi}}\bra{\beta e^{i\phi}}\bra{\beta e^{i\phi}}\\
         &=\sum_{n=0}^{\infty}P_{n}\ket{n,+}\bra{n,+},\\
\end{aligned}
\end{equation}
where the fock state is defined as
\begin{equation}
\ket{n,+}=\frac{1}{\sqrt{2^{n}n!}}(a^{\dagger}+b^{\dagger})^{n}\ket{00}_{AB},
\end{equation}
and the probability is given by
\begin{equation}
P_{n}=e^{-2\mu}\frac{(2\mu)^{n}}{n!},
\end{equation}
where $\mu=|\beta|^{2}$ is the light intensity. In the case of $\beta_{a}=\beta_{b}=\beta$ and $\phi_{a}=\phi_{b}+\pi$(mod $2\pi$)$=\phi$, the composite state of Alice and Bob can be written as
\begin{equation}
\label{constext}
\begin{aligned}
\rho_{AB}&=\frac{1}{2\pi}\int_{0}^{2\pi}d\phi\ket{\beta e^{i\phi}}\ket{-\beta e^{i\phi}}\bra{\beta e^{i\phi}}\bra{-\beta e^{i\phi}}\\
         &=\sum_{n=0}^{\infty}P_{n}\ket{n,-}\bra{n,-},\\
\end{aligned}
\end{equation}
where the fock state is defined as
\begin{equation}
\ket{n,-}=\frac{1}{\sqrt{2^{n}n!}}(a^{\dagger}-b^{\dagger})^{n}\ket{00}_{AB},
\end{equation}
with probability $P_{n}$.

In what follows, we concentrate on bounding Eve's Holevo information. Eve's general collective attack can be given by 
\begin{equation}
\label{constext}
\begin{aligned}
&U_{Eve}\ket{n,\pm}_{AB}\ket{e}_{E}=\\
&\sqrt{Y^{L}_{n,\pm}}\ket{\gamma^{L}_{n,\pm}}\ket{L}+\sqrt{Y^{R}_{n,\pm}}\ket{\gamma^{R}_{n,\pm}}\ket{R}+\sqrt{Y^{N}_{n,\pm}}\ket{\gamma^{N}_{n,\pm}}\ket{N},\\
\end{aligned}
\end{equation}
where state $\ket{e}_{E}$ is Eve's ancilla. Then, Eve is supposed to announce one of legal outcomes "Only detector $L$ clicks" "Only detector $R$ clicks" and "No detectors click" determined by her measurement results "$\ket{L}$,$\ket{R}$,$\ket{N}$", respectively. In the case of $\beta_{a}=\beta_{b}$ and $\phi_{a}=\phi_{b}$, $\ket{\gamma^{L}_{n,+}}$, $\ket{\gamma^{R}_{n,+}}$ and $\ket{\gamma^{N}_{n,+}}$ are some arbitrary quantum states referring to Eve's measurement results "$\ket{L}$,$\ket{R}$,$\ket{N}$" respectively. $Y^{L}_{n,+}$, $Y^{R}_{n,+}$ and $Y^{N}_{n,+}$ satisfying $Y^{L}_{n,+}+Y^{R}_{n,+}$+$Y^{N}_{n,+}=1$ are the yields referring to Eve's measurement results "$\ket{L}$,$\ket{R}$,$\ket{N}$" respectively. Similarly, in the case of $\beta_{a}=\beta_{b}$ and $|\phi_{a}-\phi_{b}|=\pi$, $\ket{\gamma^{L}_{n,-}}$, $\ket{\gamma^{R}_{n,-}}$ and $\ket{\gamma^{N}_{n,-}}$ are some arbitrary quantum states referring to Eve's measurement results "$\ket{L}$,$\ket{R}$,$\ket{N}$" respectively. $Y^{L}_{n,-}$, $Y^{R}_{n,-}$ and $Y^{N}_{n,-}$ satisfying $Y^{L}_{n,-}+Y^{R}_{n,-}$+$Y^{N}_{n,-}=1$ are the yields referring to Eve's measurement results "$\ket{L}$,$\ket{R}$,$\ket{N}$" respectively. 

Without loss of generality, we firstly consider the secret key rate when her measurement result is "$\ket{L}$".  When Alice and Bob both select the code mode, the initial prepared state $\ket{\alpha e^{i(k_{a}+\frac{x}{M})\pi}}$ and $\ket{\alpha e^{i(k_{b}+\frac{y}{M})\pi}}$, with matched-basis trials  $x=y$, can be given by
\begin{equation}
\label{constext}
\begin{aligned}
&\ket{\alpha e^{i\frac{x}{M}\pi}}\ket{\alpha e^{i\frac{x}{M}\pi}}=\sum_{n=0}^{\infty}\sqrt{P_{n}}e^{i\frac{nx\pi}{M}}\ket{n,+},k_{a}=k_{b}=0\\
&\ket{-\alpha e^{i\frac{x}{M}\pi}}\ket{-\alpha e^{i\frac{x}{M}\pi}}=\sum_{n=0}^{\infty}\sqrt{P_{n}}e^{i\frac{n(M+x)\pi}{M}}\ket{n,+},k_{a}=k_{b}=1\\
&\ket{\alpha e^{i\frac{x}{M}\pi}}\ket{-\alpha e^{i\frac{x}{M}\pi}}=\sum_{n=0}^{\infty}\sqrt{P_{n}}e^{i\frac{nx\pi}{M}}\ket{n,-},k_{a}=0,k_{b}=1\\
&\ket{-\alpha e^{i\frac{x}{M}\pi}}\ket{\alpha e^{i\frac{x}{M}\pi}}=\sum_{n=0}^{\infty}\sqrt{P_{n}}e^{i\frac{n(M+x)\pi}{M}}\ket{n,-},k_{a}=1,k_{b}=0.\\
\end{aligned}
\end{equation}
For the sake of simplicity, we define unnormalized states
\begin{equation}
\ket{\psi^{L/R}_{j,\pm}}=\sum_{n=0}^{\infty}\sqrt{P_{2Mn+j}Y^{L/R}_{2Mn+j,\pm}}\ket{\gamma^{L/R}_{2Mn+j,\pm}},
\end{equation}
where $j\in{\{0,1,2,\ldots,2M-1\}}$. We also define other unnormalized states
\begin{equation}
\label{constext}
\begin{aligned}
&\ket{\psi^{L/R}_{ex,\pm}}=\sum_{j=0}^{M-1}e^{i\frac{2jx\pi}{M}}\ket{\psi^{L/R}_{2j,\pm}}\\
&\ket{\psi^{L/R}_{ox,\pm}}=\sum_{j=0}^{M-1}e^{i\frac{(2j+1)x\pi}{M}}\ket{\psi^{L/R}_{2j+1,\pm}},\\
\end{aligned}
\end{equation}
After Eve's attack according to Eq.(7) and her announcing "$\ket{L}$", Alice and Bob keep trials only if $x=y$. Thus, the unnormalized state of Eve conditioned on Alice's classical bit can be given by 
\begin{equation}
\label{constext}
\begin{aligned}
&\rho^{L}_{AEx}=\frac{1}{4}\ket{0}_{A}\bra{0}\otimes(P\{\ket{\psi^{L}_{ex,+}}+\ket{\psi^{L}_{ox,+}}\}\\
              &+P\{\ket{\psi^{L}_{ex,-}}+\ket{\psi^{L}_{ox,-}}\})+\frac{1}{4}\ket{1}_{A}\bra{1}\\
              &\otimes(P\{\ket{\psi^{L}_{ex,+}}-\ket{\psi^{L}_{ox,+}}\}+P\{\ket{\psi^{L}_{ex,-}}-\ket{\psi^{L}_{ox,-}}\}),\\
\end{aligned}
\end{equation}
where $P\{\ket{x}\}=\ket{x}\bra{x}$. The probability of Alice obtaining a shifted key $(x=y)$ in a code mode when Eve announces "$\ket{L}$" is 
\begin{equation}
Q^{L}_{x}=\frac{1}{2}(|\ket{\psi^{L}_{ex,+}}|^{2}+|\ket{\psi^{L}_{ox,+}}|^{2}+|\ket{\psi^{L}_{ex,-}}|^{2}+|\ket{\psi^{L}_{ox,-}}|^{2}),
\end{equation}
and correspondingly an error click occurs if $k_{a}\oplus k_{b}=1$, thus, the error rate of shifted key $(x=y)$ is given by 
\begin{equation}
\label{constext}
\begin{aligned}
e^{L}_{x}&=\frac{|\ket{\psi^{L}_{ex,-}}|^{2}+|\ket{\psi^{L}_{ox,-}}|^{2}}{|\ket{\psi^{L}_{ex,+}}|^{2}+|\ket{\psi^{L}_{ox,+}}|^{2}+|\ket{\psi^{L}_{ex,-}}|^{2}+|\ket{\psi^{L}_{ox,-}}|^{2}}\\
         &=\frac{|\ket{\psi^{L}_{ex,-}}|^{2}+|\ket{\psi^{L}_{ox,-}}|^{2}}{2Q^{L}_{x}},\\
\end{aligned}
\end{equation}
Thanks to the strong subadditivity of von Neumann entropy (the detailed derivation how we apply the strong subadditivity is in the Appendix A of \cite{wang2018security}), Eve's Holevo information with her announcing "$\ket{L}$" is given by
\begin{equation}
\label{constext}
\begin{aligned}
&I_{AEx}^{L}\\
&\le(1-e^{L}_{x})H(\frac{|\ket{\psi^{L}_{ex,+}}|^{2}}{2(1-e^{L}_{x})Q^{L}_{x}})+ e^{L}_{x}H(\frac{|\ket{\psi^{L}_{ex,-}}|^{2}}{2e^{L}_{x}Q^{L}_{x}})\\
&\le H(\frac{|\ket{\psi^{L}_{ex,+}}|^{2}+|\ket{\psi^{L}_{ex,-}}|^{2}}{2Q^{L}_{x}}),\\
\end{aligned}
\end{equation}
where $H(x)=-x\log_2{x}-(1-x)\log_2({1-x})$ is binary Shannon entropy and the second inequality holds due to Jensen's inequality. For each trial that $x=y$ and Eve announces "$\ket{L}$", the secret key rate is given by 
\begin{equation}
R_{x}^{L}=Q^{L}_{x}(1-fH(e^{L}_{x})-I_{AEx}^{L}),
\end{equation}
where $f$ is error correction efficiency. What we need to do next is to calculate the average secret key rate for different $x$ when Eve announces "$\ket{L}$". Without considering the sifting factor, the average secret key rate when Eve announces "$\ket{L}$" is given by
\begin{equation}
R^{L}=\frac{1}{M}\sum_{x=0}^{M-1}R_{x}^{L}=\frac{1}{M}\sum_{x=0}^{M-1}Q^{L}_{x}(1-fH(e^{L}_{x})-I_{AEx}^{L}),
\end{equation}
We use $Q^{L}$ to denote the average gain and $e^{L}$ to denote the average error rate of shifted key, which are written as
\begin{equation}
\label{constext}
\begin{aligned}
&Q^{L}=\frac{1}{M}\sum_{x=0}^{M-1}Q^{L}_{x}\\
&e^{L}=\frac{\sum_{x=0}^{M-1}Q^{L}_{x}e^{L}_{x}}{\sum_{x=0}^{M-1}Q^{L}_{x}},\\
\end{aligned}
\end{equation}
Thanks to the concavity of binary Shannon entropy, we utilize Jensen's inequality to minimize $R^{L}$. For the second term of Eq.(16) on the right, we have
\begin{equation}
\frac{1}{M}\sum_{x=0}^{M-1}Q^{L}_{x}H(e^{L}_{x})\le Q^{L}H(\frac{\frac{1}{M}\sum_{x=0}^{M-1}Q^{L}_{x}e^{L}_{x}}{Q^{L}})=Q^{L}H(e^{L}),
\end{equation}
The condition for equality of Eq.(18) is that $e^{L}_{0}=e^{L}_{1}=\ldots=e^{L}_{M-1}$. Similarly, for the third term of Eq.(16) on the right, we have
\begin{equation}
\label{constext}
\begin{aligned}
&\frac{1}{M}\sum_{x=0}^{M-1}Q^{L}_{x}I_{AEx}^{L}\\
\le&\frac{1}{M}\sum_{x=0}^{M-1}Q^{L}_{x}H(\frac{|\ket{\psi^{L}_{ex,+}}|^{2}+|\ket{\psi^{L}_{ex,-}}|^{2}}{2Q^{L}_{x}})\\
\le&Q^{L}H(\frac{1}{2MQ^{L}}\sum_{x=0}^{M-1}|\sum_{j=0}^{M-1}e^{i\frac{2jx\pi}{M}}\ket{\psi^{L}_{2M+2j,+}}|^{2}\\
                                         &+|\sum_{j=0}^{M-1}e^{i\frac{2jx\pi}{M}}\ket{\psi^{L}_{2M+2j,-}}|^{2})\\
=&Q^{L}H(\frac{\sum_{j=0}^{M-1}|\ket{\psi^{L}_{2M+2j,+}}|^{2}+|\ket{\psi^{L}_{2M+2j,-}}|^{2}}{2Q^{L}}).\\
\end{aligned}
\end{equation}
Here we define $I_{AE}^{L}=H(\frac{\sum_{j=0}^{M-1}|\ket{\psi^{L}_{2M+2j,+}}|^{2}+|\ket{\psi^{L}_{2M+2j,-}}|^{2}}{2Q^{L}})$. Consequently, we have
\begin{equation}
R^{L}\ge Q^{L}(1-fH(e^{L})-I_{AE}^{L}).
\end{equation}

Similarly, when Eve's measurement result is "$\ket{R}$", the analysis of secret key rate is almost the same with the ones when she announces "$\ket{L}$". Thus, the secret key rate when Eve announces "$\ket{R}$" is given by
\begin{equation}
R^{R}\ge Q^{R}(1-fH(e^{R})-I_{AE}^{R}),
\end{equation}
where $I_{AE}^{R}$ is given by
\begin{equation}
\label{constext}
\begin{aligned}
I_{AE}^{R}=H(\frac{\sum_{j=0}^{M-1}|\ket{\psi^{R}_{2M+2j,-}}|^{2}+|\ket{\psi^{R}_{2M+2j,+}}|^{2}}{2Q^{R}}).
\end{aligned}
\end{equation}

The trials when Eve's measurement result is "$\ket{N}$" will not contribute to the secret key. Thus, the total secret key rate is $R=R^{L}+R^{R}$. The total gain and the total error rate of shifted key are given by
\begin{equation}
\label{constext}
\begin{aligned}
&Q=Q^{L}+Q^{R}\\
&e=\frac{Q^{L}e^{L}+Q^{R}e^{R}}{Q},\\
\end{aligned}
\end{equation}
In order to find the lower bound of the total secret key rate $R$, we apply the Jensen's inequality to the estimation items in Eq.(15) and Eq. (21), and we can get
\begin{equation}
Q^{L}H(e^{L})+Q^{R}H(e^{R})\le QH(\frac{Q^{L}e^{L}+Q^{R}e^{R}}{Q})=QH(e),
\end{equation}
where the equality holds when $e^{L}=e^{R}=e$, and
\begin{equation}
\label{constext}
\begin{aligned}
&Q^{L}I_{AE}^{L}+Q^{R}I_{AE}^{R}\\
\le&Q[H(\frac{1}{2Q}{\sum_{j=0}^{M-1}|\ket{\psi^{L}_{2M+2j,+}}|^{2}+|\ket{\psi^{R}_{2M+2j,-}}|^{2}}\\
                       &+|\ket{\psi^{L}_{2M+2j,-}}|^{2}+|\ket{\psi^{R}_{2M+2j,+}}|^{2})],\\
\end{aligned}
\end{equation}
where we define 
\begin{equation}
\label{constext}
\begin{aligned}
I_{AE}&=H(\frac{1}{2Q}{\sum_{j=0}^{M-1}|\ket{\psi^{L}_{2M+2j,+}}|^{2}+|\ket{\psi^{R}_{2M+2j,-}}|^{2}}\\
                                                               &+|\ket{\psi^{L}_{2M+2j,-}}|^{2}+|\ket{\psi^{R}_{2M+2j,+}}|^{2}).\\
\end{aligned}
\end{equation}
Consequently, the total secret key rate formula can be expressed by 
\begin{equation}
R\ge \frac{1}{M}Q(1-fH(e)-I_{AE}),
\end{equation}
where $1/M$ is the shifting factor. 
And the problem of finding the lower bound of the total secret key rate can be converted into finding the upper bound of $I_{AE}$,
\begin{equation}
\label{constext}
\begin{aligned}
I_{AE}&\le H(\frac{1}{2Q}{\sum_{j=0}^{M-1}|\ket{\psi^{L}_{2M+2j,+}}|^{2}+|\ket{\psi^{R}_{2M+2j,-}}|^{2}}\\
                       &+|\ket{\psi^{L}_{2M+2j,-}}|^{2}+|\ket{\psi^{R}_{2M+2j,+}}|^{2})\\
      &\text{ with constrants}\\
0\le&|\ket{\psi^{L/R}_{2M+j,\pm}}|^{2}\le |\sum_{n=0}^{\infty}\sqrt{P_{2Mn+j}Y^{L/R}_{2Mn+j,\pm}}|^{2}\\
&{\sum_{j=0}^{M-1}|\ket{\psi^{L}_{2M+2j,+}}|^{2}+|\ket{\psi^{R}_{2M+2j,-}}|^{2}}\\
                &+|\ket{\psi^{L}_{2M+2j,-}}|^{2}+|\ket{\psi^{R}_{2M+2j,+}}|^{2}\le Q.\\
\end{aligned}
\end{equation}

\hfill

\noindent{\bf Simulation}

\noindent In this section, we simulate the performance of our TF-QKD protocols, and the simulation method is very similar to Ma et al. \cite{ma2018phase}. Ideally, for Protocol \uppercase\expandafter{\romannumeral1}, Alice and Bob can estimate $Y^{L/R}_{n,\pm}$ precisely by infinite decoy-state method.

We assume that the total efficiency of channels and detectors is $\eta$, dark counting rate of single photon detectors (SPD) is $d$ per trial, the optical misalignment is $e_{mis}$, and the mean photon number of each pulse emitted by Alice and Bob is $\mu$. The counting rate is given by
\begin{equation}
\begin{aligned}
Q&=(1-d)(1-e^{-2\eta\mu})+2d(1-d)e^{-2\eta\mu}\\
 &=(1-d)(1-e^{-2\eta\mu}+2de^{-2\eta\mu}),\\
\end{aligned}
\end{equation}
and the error rate is
\begin{equation}
e=\frac{(1-d)[e_{mis}-(e_{mis}-d)e^{-2\eta\mu}]}{Q}.
\end{equation}
Applying infinite decoy states, $Y^{L/R}_{n,\pm}$ can be given by
\begin{equation}
\begin{aligned}
&Y^{L}_{n,+}=Y^{R}_{n,-}=(1-d)[1-e_{mis}-(1-e_{mis}-d)(1-\eta)^{n}]\\
&Y^{L}_{n,-}=Y^{R}_{n,+}=(1-d)[e_{mis}-(e_{mis}-d)(1-\eta)^{n}].\\
\end{aligned}
\end{equation}
We define
\begin{equation}
\begin{aligned}
&Y^{L}_{n,+}=Y^{R}_{n,-}=Y^{c}_{n}\\
&Y^{L}_{n,-}=Y^{R}_{n,+}=Y^{e}_{n}\\
&Y_{n}=Y^{c}_{n}+Y^{e}_{n}=(1-d)[1-(1-2d)(1-\eta)^{n}]\\
&X^{c}_{2M+j}=\frac{|\ket{\psi^{L}_{2M+j,+}}|^{2}+|\ket{\psi^{R}_{2M+j,-}}|^{2}}{2}\\
&X^{e}_{2M+j}=\frac{|\ket{\psi^{L}_{2M+j,-}}|^{2}+|\ket{\psi^{R}_{2M+j,+}}|^{2}}{2}\\
&X_{2M+j}=X^{c}_{2M+j}+X^{e}_{2M+j}.\\
\end{aligned}
\end{equation}
Thanks to Cauchy inequality, we have
\begin{equation}
\begin{aligned}
    &(\sum_{n=0}^{\infty}\sqrt{P_{n}Y^{c}_{n}})^{2}+(\sum_{n=0}^{\infty}\sqrt{P_{n}Y^{e}_{n}})^{2}\\
=   &\sum_{n=0}^{\infty}P_{n}(Y^{c}_{n}+Y^{e}_{n})+\sum_{n\ne n^{'}}^{\infty}\sqrt{P_{n}P_{n^{'}}}(\sqrt{Y^{c}_{n}Y^{c}_{n^{'}}}+\sqrt{Y^{e}_{n}Y^{e}_{n^{'}}})\\
\le &\sum_{n=0}^{\infty}P_{n}Y_{n}+\sum_{n\ne n^{'}}^{\infty}\sqrt{P_{n}P_{n^{'}}Y_{n}Y_{n^{'}}}\\
=   &(\sum_{n=0}^{\infty}\sqrt{P_{n}Y_{n}})^{2}.\\
\end{aligned}
\end{equation}
Thus, we can get an equivalent upper bound of $I_{AE}$ given by
\begin{equation}
\label{constext}
\begin{aligned}
I_{AE}&\le H(\frac{\sum_{j=0}^{M-1}X_{2M+2j}}{Q})\\
      &\text{ with constraints}\\
0\le&X_{2M+2j}\le (\sum_{n=0}^{\infty}\sqrt{P_{2Mn+2j}Y_{2Mn+2j}})^{2}\\
&\sum_{j=0}^{M-1}X_{2M+2j}\le \frac{Q}{2}.\\
\end{aligned}
\end{equation}

\hfill

\noindent{\bf Security proof for Protocol \uppercase\expandafter{\romannumeral2}}

\noindent The security proof of Protocol \uppercase\expandafter{\romannumeral2} is almost the same as Protocol \uppercase\expandafter{\romannumeral1}. In Protocol \uppercase\expandafter{\romannumeral2}, Eve's general collective attack is given by
\begin{equation}
\label{constext}
\begin{aligned}
&U_{Eve}\ket{l,k}_{AB}\ket{e}_{E}=\\
&\sqrt{Y^{L}_{l,k}}\ket{\gamma^{L}_{l,k}}\ket{L}+\sqrt{Y^{R}_{l,k}}\ket{\gamma^{R}_{l,k}}\ket{R}+\sqrt{Y^{N}_{l,k}}\ket{\gamma^{N}_{l,k}}\ket{N}.\\
\end{aligned}
\end{equation} 
where $\ket{l,k}_{AB}$ represents the photon-number base prepared by Alice and Bob, $\ket{\gamma^{L}_{l,k}}$, $\ket{\gamma^{R}_{l,k}}$ and $\ket{\gamma^{N}_{l,k}}$ are some arbitrary quantum states referring to Eve's measurement results "$\ket{L}$,$\ket{R}$,$\ket{N}$" respectively.Besides, $Y^{L}_{l,k}$, $Y^{R}_{l,k}$ and $Y^{N}_{l,k}$ satisfying $Y^{L}_{l,k}+Y^{R}_{l,k}$+$Y^{N}_{l,k}=1$ are the yields referring to Eve's measurement results "$\ket{L}$,$\ket{R}$,$\ket{N}$" respectively. Compared to the expression of Eve's general collective attack in Protocol \uppercase\expandafter{\romannumeral1}, it can be argued that the general collective attack is actually the same as Protocol \uppercase\expandafter{\romannumeral1} if we set
\begin{equation}
\sqrt{P_{n}Y^{L/R}_{n,\pm}}\ket{\gamma^{L/R}_{n,\pm}}= \sum_{l=0,l+k=n}^{n}(\pm1)^{l}\sqrt{P_{l,k}Y^{L/R}_{l,k}}\ket{\gamma^{L/R}_{l,k}}.
\end{equation}
Consequently, applying the security proof method to Protocol \uppercase\expandafter{\romannumeral2}, we find that the expression of the upper bound of $I_{AE}$ is same as the one of Protocol \uppercase\expandafter{\romannumeral1}. In Protocol \uppercase\expandafter{\romannumeral2}, for removing phase post-selection, we estimate the yield $Y_{l,k}$ rather than $Y_{n}$ to bound $X_{2M+2j}$. Combining Eq.(9) and Eq.(36), we obtain

\begin{equation}
\label{constext}
\begin{aligned}
  &\frac{1}{2}(|\ket{\psi^{L/R}_{2M+2j,+}}|^{2}+|\ket{\psi^{L/R}_{2M+2j,-}}|^{2})\\
=&\frac{1}{2}(|\sum_{n=0}^{\infty}\sqrt{P_{2Mn+2j}Y^{L/R}_{2Mn+2j,+}}\ket{\gamma^{L/R}_{2Mn+2j,+}}|^{2}\\
+&|\sum_{n=0}^{\infty}\sqrt{P_{2Mn+2j}Y^{L/R}_{2Mn+2j,-}}\ket{\gamma^{L/R}_{2Mn+2j,-}}|^{2})\\
=&\frac{1}{2}(|\sum_{n=0}^{\infty}\sum_{l=0}^{l+k=2Mn+2j}\sqrt{P_{l,k}Y^{L/R}_{l,k}}\ket{\gamma^{L/R}_{l,k}}|^{2}\\
+&|\sum_{n=0}^{\infty}\sum_{l=0}^{l+k=2Mn+2j}(-1)^{l}\sqrt{P_{l,k}Y^{L/R}_{l,k}}\ket{\gamma^{L/R}_{l,k}}|^{2})\\
=&|\sum_{n=0}^{\infty}\sum_{l,k\in even}^{l+k=2Mn+2j}\sqrt{P_{l,k}Y^{L/R}_{l,k}}\ket{\gamma^{L/R}_{l,k}}|^{2}\\
+&|\sum_{n=0}^{\infty}\sum_{l,k\in odd}^{l+k=2Mn+2j}\sqrt{P_{l,k}Y^{L/R}_{l,k}}\ket{\gamma^{L/R}_{l,k}}|^{2}\\
\le&(\sum_{n=0}^{\infty}\sum_{l,k\in even}^{l+k=2Mn+2j}\sqrt{P_{l,k}Y^{L/R}_{l,k}})^{2}\\
+&(\sum_{n=0}^{\infty}\sum_{l,k\in odd}^{l+k=2Mn+2j}\sqrt{P_{l,k}Y^{L/R}_{l,k}})^{2}.\\
\end{aligned}
\end{equation}
where $even$ and $odd$ are the assembles referring to even number set and odd number set respectively. Similar to Eq.(33), by utilizing Cauchy inequality, we have
\begin{equation}
\label{constext}
\begin{aligned}
   &(\sum_{n=0}^{\infty}\sum_{l,k}^{l+k=2Mn+2j} \sqrt{P_{l,k}Y^{L}_{l,k}})^{2}\\
 + &(\sum_{n=0}^{\infty}\sum_{l,k}^{l+k=2Mn+2j} \sqrt{P_{l,k}Y^{R}_{l,k}})^{2}\\
\le&(\sum_{n=0}^{\infty}\sum_{l,k}^{l+k=2Mn+2j} \sqrt{P_{l,k}Y_{l,k}})^{2}, \\
\end{aligned}
\end{equation}
where $Y_{l,k}=Y^{L}_{l,k}+Y^{R}_{l,k}$. Due to the decoy-state method implemented, $Y_{l,k}$ satisfies the constraints  
\begin{equation}
Q^{\mu_{a}\mu_{b}}=\sum_{l,k}P^{\mu_{a}\mu_{b}}_{l,k}Y_{l,k}.
\end{equation}
Thus, we have obtained the upper bound of $X_{2M+2j}$ given as follows
\begin{equation}
\label{constext}
\begin{aligned}
X_{2M+2j}\le&(\sum_{n=0}^{\infty}\sum_{l,k\in even}^{l+k=2Mn+2j}\sqrt{P_{l,k}Y_{l,k}})^{2}\\
+&(\sum_{n=0}^{\infty}\sum_{l,k\in odd}^{l+k=2Mn+2j}\sqrt{P_{l,k}Y_{l,k}})^{2}.\\
\end{aligned}
\end{equation}
Briefly, the upper bound of $I_{AE}$ in Protocol \uppercase\expandafter{\romannumeral2} is given by,
\begin{equation}
\label{constext}
\begin{aligned}
I_{AE}&\le H(\frac{\sum_{j=0}^{M-1}X_{2M+2j}}{Q})\\
      &\text{ with constriants}\\
0\le X_{2M+2j}\le& (\sum_{n=0}^{\infty}\sum_{l,k\in even}^{l+k=2Mn+2j}\sqrt{P_{l,k}Y_{l,k}})^{2}\\
+&(\sum_{n=0}^{\infty}\sum_{l,k\in odd}^{l+k=2Mn+2j}\sqrt{P_{l,k}Y_{l,k}})^{2}\\
&\sum_{j=0}^{M-1}X_{2M+2j}\le \frac{Q}{2}.\\
\end{aligned}
\end{equation}

\hfill

\noindent{\bf $I_{AE}$ as a function of $M$}

\noindent For the sake of analyzing the upper bound of $I_{AE}$ with the increase of $M$, we define the upper bound of $\sum_{j=0}^{M-1}X_{2M+2j}$ as a function of positive integer $M$, which is given by
\begin{equation}
F(M)=\sum_{j=0}^{M-1}(\sum_{n=0}^{\infty}\sqrt{P_{2Mn+2j}Y_{2Mn+2j}})^{2}.
\end{equation}
As binary Shannon entropy $H(x)$ increases when $0\le x\le 1/2$ and decreases when $1/2\le x\le 1$, it's sufficient to consider the case of $F(M)\le Q/2$. It can be proven that
\begin{equation}
F(1)\ge F(M)\ge F(NM)\ge F(\infty) ,
\end{equation}
where $N$ is a positive integer. In order to prove Eq.(43), we rewrite Eq.(42) as follows
\begin{equation}
G(M)=\sum_{j=0}^{M-1}(\sum_{n=0}^{\infty}A_{2Mn+2j})^{2}
\end{equation}
where we denote $F(M)$ and $\sqrt{P_{2Mn+2j}Y_{2Mn+2j}}$ as $G(M)$ and $A_{2Mn+2j}$ respectively. For $A_{2Mn+2j}$ is absolutely a nonnegative term, we have 
\begin{equation}
\label{constext}
\begin{aligned}
G(1)&=(\sum_{j=0}^{\infty}A_{2j})^{2}=(\sum_{n=0}^{\infty}A_{2n})^{2}\\
    &=(\sum_{j=0}^{M-1}(\sum_{n=0}^{\infty}A_{2Mn+2j}))^{2}\\
    &=\sum_{j=0}^{M-1}(\sum_{n=0}^{\infty}A_{2Mn+2j})^{2}\\
    &+\sum_{j\neq j^{'}}^{M-1}(\sum_{n=0}^{\infty}A_{2Mn+2j})(\sum_{n=0}^{\infty}A_{2Mn+2j^{'}})\\
    &\ge \sum_{j=0}^{M-1}(\sum_{n=0}^{\infty}A_{2Mn+2j})^{2}\\
    &=G(M).\\
\end{aligned}
\end{equation}
where the inequality holds because of the nonnegative cross term $\sum_{j\neq j^{'}}^{M-1}(\sum_{n=0}^{\infty}A_{2Mn+2j})(\sum_{n=0}^{\infty}A_{2Mn+2j^{'}})$. Similarly, 
\begin{equation}
\label{constext}
\begin{aligned}
G(M)&=\sum_{j=0}^{M-1}(\sum_{n=0}^{\infty}A_{2Mn+2j})^{2}\\
    &=\sum_{j=0}^{M-1}(\sum_{j^{'}=0}^{N-1}\sum_{n=0}^{\infty}A_{2N(Mn+j)+2j^{'}})^{2}\\
    &\ge \sum_{j=0}^{M-1}\sum_{j^{'}=0}^{N-1}(\sum_{n=0}^{\infty}A_{2N(Mn+j)+2j^{'}})^{2}\\
    &=\sum_{j=0}^{M-1}\sum_{j^{'}=0}^{N-1}(\sum_{n=0}^{\infty}A_{2NMn+2(Nj+j^{'})})^{2}\\
    &=\sum_{k=0}^{NM-1}(\sum_{n=0}^{\infty}A_{2NMn+2k})^{2}\\
    &=G(NM),\\
\end{aligned}
\end{equation}
where we use subscript $k$ instead of $Nj+j^{'}$. The nonnegative cross term vanishes when $M\to \infty$, then, we have 
\begin{equation}
G(\infty)=\sum_{n=0}^{\infty}A_{2n}^{2}.
\end{equation}
Thus, we have proven Eq.(43). Then we obtain that the upper bound of $I_{AE}$ decreases with $M$ exponentially increasing. In other words, the achievable distance becomes longer as $M$ exponentially increasing. As a result, the achievable distance comes to a limitation when $M$ tends to infinity.

\hfill

\noindent{\bf Finite-decoy method for Protocol \uppercase\expandafter{\romannumeral2} with $M=2$}

For Protocol \uppercase\expandafter{\romannumeral2} does not require phase post-selection in the test mode, it is more practical than Protocol \uppercase\expandafter{\romannumeral1}. For Protocol \uppercase\expandafter{\romannumeral2}, it almost reaches the limitary transmission distance with $M=2$ shown in TABLE \uppercase\expandafter{\romannumeral3}, thus, it is interesting and necessary to consider applying finite decoy states in the test mode. 

When finite decoy states are implemented, finding the upper bound of $I_{AE}$ is equivalent to the following optimized problem
\begin{equation}
\label{constext}
\begin{aligned}
&\text{Max}: \\
&\sum_{j=0}^{M-1}(\sum_{n=0}^{\infty}\sum_{l,k\in even}^{l+k=2Mn+2j}\sqrt{P_{l,k}Y_{l,k}})^{2} \\
                           &+(\sum_{n=0}^{\infty}\sum_{l,k\in odd}^{l+k=2Mn+2j}\sqrt{P_{l,k}Y_{l,k}})^{2}\\
& \text{ s.t. }\\   
&\sum_{l,k=0}^{6}P^{\mu_{a}\mu_{b}}_{l,k}Y_{l,k}\le Q^{\mu_{a}\mu_{b}}\le \sum_{l,k=0}^{6}P^{\mu_{a}\mu_{b}}_{l,k}Y_{l,k}+1-\sum_{l,k=0}^{6}P^{\mu_{a}\mu_{b}}_{l,k}\\
& \text{ and }  \\
& \sum_{j=0}^{M-1}(\sum_{n=0}^{\infty}\sum_{l,k\in even}^{l+k=2Mn+2j}\sqrt{P_{l,k}Y_{l,k}})^{2} \\
                           &+(\sum_{n=0}^{\infty}\sum_{l,k\in odd}^{l+k=2Mn+2j}\sqrt{P_{l,k}Y_{l,k}})^{2} \le \frac{Q}{2}.\\
\end{aligned}
\end{equation}
where $\mu_{a},\mu_{b}\in \{\mu_{1},\mu_{2},\mu_{3}\}$. As Fig 4 shows, the performance is maintained using only three intensity settings. That is, we only need three decoy intensities $\{\mu_{1},\mu_{2},\mu_{3}\}$, and the signal intensity is chosen from one of them.

\begin{figure}[htbp]
\centering
\includegraphics[width=\linewidth]{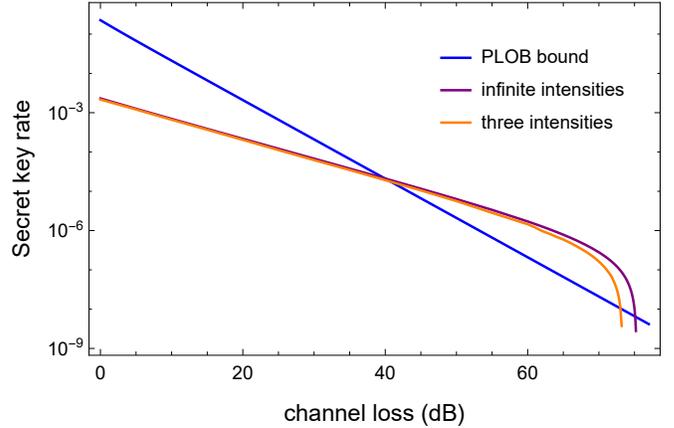}
\caption{Secret key rate R versus channel loss for Protocol \uppercase\expandafter{\romannumeral2} with $M=2$ ($M$ is the number of random phases) : The curves represent the secure key rate in the case of infinite intensities, three intensities and the Pirandola-Laurenza-Ottaviani-Banchi (PLOB) bound respectively.}
\label{fig:false-color}
\end{figure}

\hfill

\noindent {\bf Data availability}

\noindent The data that support the findings of this study are available from the corresponding author upon reasonable request

\hfill

\noindent{\bf ACKNOWLEDGEMENTS}

\noindent This work has been supported by the National Key Research and Development Program of China (Grant No. 2018YFA0306400); National Natural Science Foundation of China (Grant Nos. 61822115, 61961136004, 61775207, 61702469, 61771439, 61627820, 61675189); National Cryptography Development Fund (Grant No. MMJJ20170120); Anhui Initiative in Quantum Information Technologies.

\hfill

\noindent{\bf Competing interests}

\noindent The authors declare that they have no competing interests.

\hfill

\noindent{\bf AUTHOR CONTRIBUTIONS}

\noindent Z.-Q.Y., S.W., W.C., G.-C.G., Z.-F.H., and R.W. conceived the basic idea of the security proof. R.W. finished the details of the security proof. Z.-Q.Y., R.W., F.-Y.L., C.-M.Z., W.H., and B.-J.X. designed the simulations. Z.-Q.Y. and R.W. wrote the paper.

\hfill

\end{document}